\documentclass{svjour3}                     % onecolumn (standard format)
\smartqed  % flush right qed marks, e.g. at end of proof
\usepackage{graphicx}
\journalname{General Relativity and Gravitation}
\begin{document}

\title{Spin and mass of the nearest supermassive black hole}
\author{Vyacheslav I. Dokuchaev}
\authorrunning{Vyacheslav I. Dokuchaev} % if too long for running head
\institute{V. Dokuchaev \at
              Institute for Nuclear Research, Russian Academy of Sciences, \\
60th October Anniversary Prospect 7a, 117312 Moscow, Russia \\
              \email{dokuchaev@inr.ac.ru}        }

\date{Received: date / Accepted: date}

\maketitle

\begin{abstract}
Quasi-Periodic Oscillations (QPOs) of the hot plasma spots or clumps orbiting an accreting
black hole contain information on the black hole mass and spin. The promising observational
signatures for the measurement of black hole mass and spin are the latitudinal oscillation
frequency of the bright spots in the accretion flow and the frequency of black hole event
horizon rotation. Both of these frequencies are independent of the accretion model and defined
completely by the properties of the black hole gravitational field. Interpretation of the known
QPO data by dint of a signal modulation from the hot spots in the accreting plasma reveals the
Kerr metric rotation parameter, $a=0.65\pm0.05$, and mass, $M=(4.2\pm0.2)10^6M_\odot$, of the
supermassive black hole in the Galactic center. At the same time, the observed 11.5 min QPO
period is identified with a period of the black hole event horizon rotation, and, respectively,
the 19 min period is identified with a latitudinal oscillation period of hot spots in the
accretion flow. The described approach is applicable to black holes with a low accretion rate,
when accreting plasma is transparent up to the event horizon region.
\end{abstract}

\section{Introduction}

In the Galactic Center dwells the nearest supermassive black hole with a mass $M = (4.1\pm0.4)\,10^6M_\odot$, measured by observations of the orbital parameters of the fast moving S0 stars \cite{Gillessen09,Meyer12}. The major problem with observations of the supermassive black hole SgrA* in the Galactic Center is that it is a `dormant' quasar, i.\,e., nearly completely inactive with a very rare splashing of activity. Nevertheless, there are serendipitous observations of (Quasi-Periodic Oscillations) QPOs from SgrA* with a sufficiently high statistical significance in the X-rays \cite{Aschenbach04} and in the near-infrared \cite{Genzel03}. It is demonstrated below the ability of the QPO observations for the measuring of both the mass and spin of the astrophysical black holes.

The usual interpretation relates QPOs with the resonances in the accretion disks
\cite{Aschenbach04,Genzel03,McClintockRemillard06,Kato10,Dolence12}. A
weak point of the resonance QPO interpretation is the ambiguity caused by the dependence on the
accretion model used. It seems that the resonance QPO models are applicable to black holes in
the Active Galactic Nuclei (AGN) and in the stellar binaries with the high accretion rates.

The others promising approaches for revealing the black hole rotation are the continuum fitting
method for the relativistic accretion disk models
\cite{NovikovThorne73,PageThorne74,McClintock11}, the modelling of the spectral lines
broadening in the accretion flow \cite{Fabian89,Brenneman06,IronLine07} and correlation between
jet power and black hole spin \cite{Narayan12}. These approaches also depend on the used
accretion models.

What is described below is the alternative QPO interpretation, related to the oscillation
frequencies of the numerous hot spots in the accretion plasma
\cite{Syunyaev73,Abramowicz92,Zakharov94,JunFukue,Broderick06,Wang12}, which are independent of the
accretion model and defined completely by the properties of the black hole gravitational field.
In this interpretation the QPOs instead of the pure periodic ones result from the simultaneous
emission of numerous clumps of hot plasma, existing in the accretion flow at different radial
distances from black hole \cite{Hawley01,Armitage01,Reynolds01}. The described approach is
directly applicable to black holes with a low accretion rate, when accreting plasma is
transparent up to the event horizon region. It is supposed that hot spots are generated in the accretion flow by plasma instability and turbulence. These hot spots will be viewed by the distant observer on the disk surface in the case of the optically thick accretion. The hot spots are directly produced in the magnetohydrodynamic simulations of accretion flow \cite{Reynolds03}.

The supermassive black hole in the Galactic Center is the most favorable case. This
supermassive black hole is activated from time to time by the episodic accretion of tidally
disrupted stars \cite{hills75,dokoz77,CarLum82,rees88,dok89,dokuch91,Genzel10}, accompanied by
the formation of shocks, jets and acceleration of cosmic rays
\cite{rees84,begblrees84,Revnivtsev04,Cheng12}. Especially crucial for the electromagnetic
extraction of energy from black holes and for the efficiency of generation and acceleration of
cosmic rays is the value of Kerr metric angular momentum \cite{BlandZna77}. The nonrotating spherically symmetric Schwarzschild black holes seems to be very exotic objects in the Universe. A typical black hole must rotate rather fast, obtaining angular momentum either from the collapsing progenitor massive star, or after the coalescence with an other black hole, or due to the spinning up by an accretion matter.

We use units with $G=c=1$ and define the spin parameter of Kerr black hole as $a=J/M^2$, where $M$ and $J$ are, respectively, the black hole mass and angular momentum.

Approaching to the extreme Kerr black hole state with $a=1$ is only possible as an infinite limiting process \cite{bardeen70}, according to the third law of the black hole (thermo)dynamics \cite{bardCarHaw73}. In the case of the astrophysically interesting thin disk accretion, the black hole will finally spin up to the ``canonical'' value of the Kerr spin parameter $a_*=0.9982$, which corresponds to the untwisting black hole in the thin disk model under the back reaction influence of the accreting flow \cite{Thorne74}.

Equations of motion for test particles (e.\,g., compact gas clouds or clumps of hot plasma) with a mass $\mu$ in the Kerr metric in the Boyer--Lindquist coordinates $(t,r,\theta,\varphi)$ are
\cite{Carter68}:
\begin{eqnarray}
 \rho^2\frac{dr}{d\lambda} &=& \pm \sqrt{V_r}, \quad
 \rho^2\frac{d\theta}{d\lambda} = \pm\sqrt{V_{\rm \theta}}, \label{rmot} \\
 \rho^2\frac{d\varphi}{d\lambda} &=& L\sin^{-2}\theta+a(\Delta^{-1}P-E),
 \label{phimot} \\
 \rho^2\frac{dt}{d\lambda} &=& a(L-aE\sin^{2}\theta)+(r^2\!+\!a^2)\Delta^{-1}P,
  \label{tmot}
\end{eqnarray}
where $\lambda=\tau/\mu$, $\tau$ --- is the proper time of a particle and
\begin{eqnarray}
 V_r &=& P^2-\Delta[\mu^2r^2+(L-aE)^2+Q], \label{Vr} \\
 V_{\rm \theta} &=& Q-\cos^2\theta[a^2(\mu^2-E^2)+L^2\sin^{-2}\theta],  \label{Vtheta} \\
 P &=& E(r^2+a^2)-a L, \\%\quad
 \rho^2 &=& r^2+a^2\cos^2\theta, \\%\quad
 \Delta &=& r^2-2r+a^2.  \label{Delta}
 \end{eqnarray}
The motion of a test particle is completely defined by three integrals of motion: the total
particle energy $E$, the azimuthal component of the angular momentum $L$ and the Carter
constant $Q$, related to the total angular momentum of the particle and non-equatorial motion. It is useful to choose the dimensionless variables and parameters: $t\Rightarrow t/M$,
$r\Rightarrow r/M$, $a\Rightarrow a/M$, $E\Rightarrow E/\mu$, $L\Rightarrow L/(M\mu)$,
$Q\Rightarrow Q/(M^2\mu^2)$. For clarity we will use the designation $x=r/M$ for the dimensionless radial distance.

The effective potentials $V_r$ and $V_{\rm \theta}$ in (\ref{Vr})
and (\ref{Vtheta}) determine the motion of particles in the radial $r$-direction and
latitudinal $\theta$-direction, respectively. The radius of the black hole event horizon is
$x_{\rm h}=1+\sqrt{1-a^2}$.

\section{Quasi-periodic plunging trajectories}

\begin{figure}[t]
\begin{center}
\includegraphics[angle=0,width=0.95\textwidth]{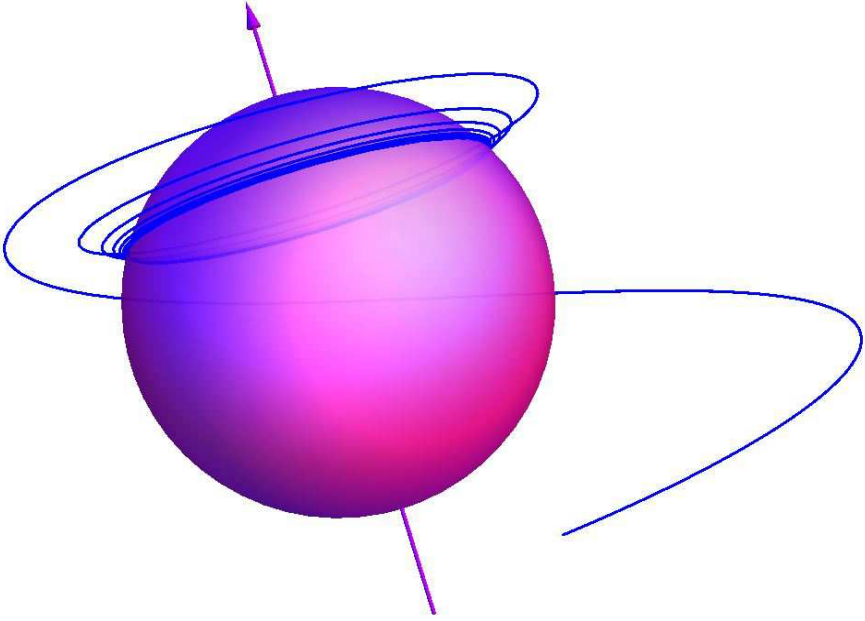}
\end{center}
\caption{The planet trajectory with a total energy $E=0.85$, azimuthal angular momentum $L=1.7$
and Carter constant $Q=1$, infalling into the black hole with a spin $a=0.9982$ and event
horizon radius $x_{\rm h}=1.063$. The trajectory is ``winding up'' with the angular velocity
$\Omega_\varphi\to\Omega_{\rm h}$, by approaching the black hole horizon at the northern
hemisphere. Trajectory is shown here and in the further similar Figures in the Boyer--Lindquist
coordinates.)} \label{fig1}
\end{figure}

\begin{figure}%[th]
\begin{center}
\includegraphics[angle=0,width=0.85\textwidth]{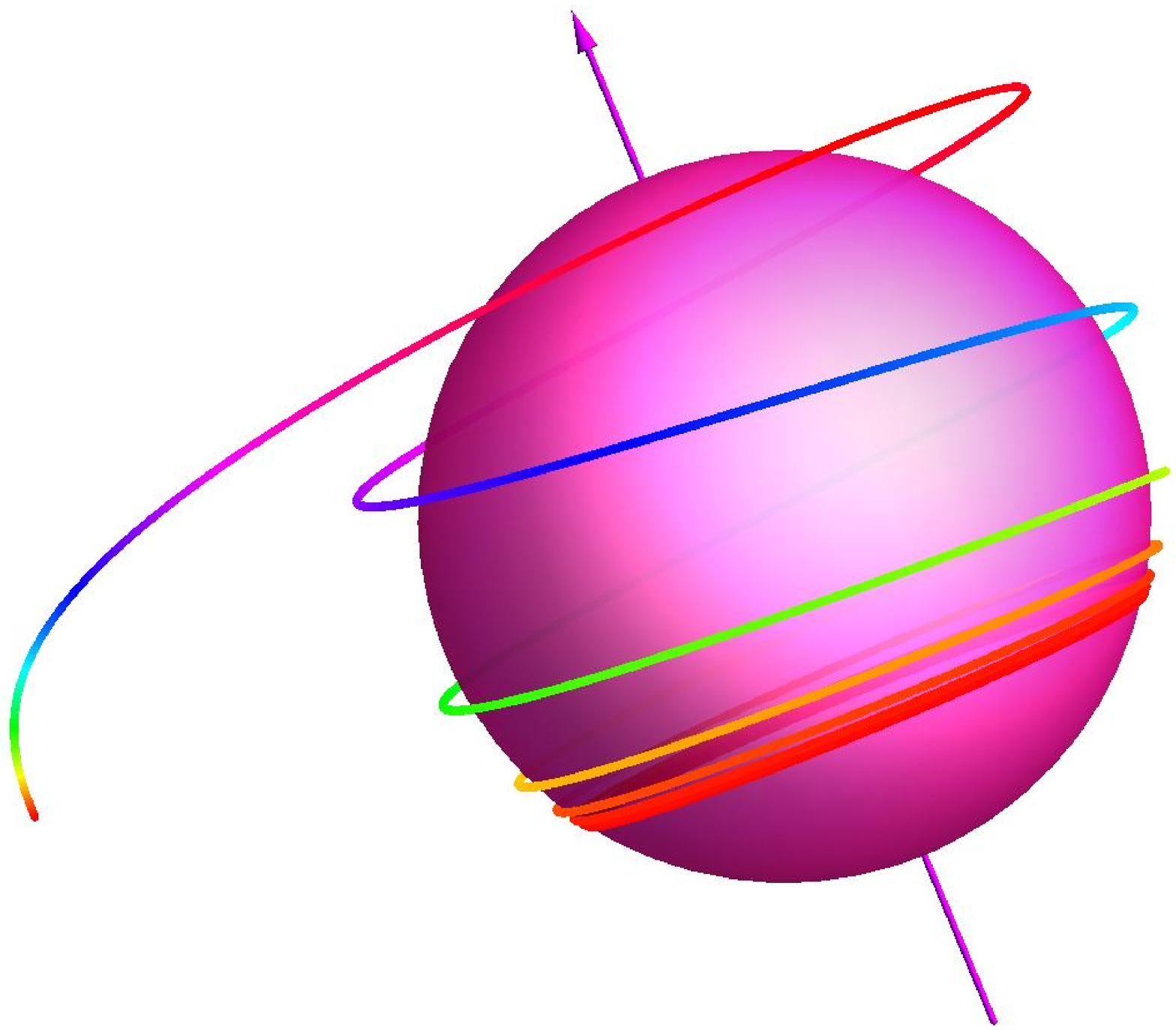}
\end{center}
\caption{The photon trajectory with an impact parameter $b=L/E=2$ and Carter constant $Q=2$,
infalling into the black hole with $a=0.9982$ and $x_{\rm h}=1.063$. The trajectory is
``winding up'' with the angular velocity $\Omega_\varphi\to\Omega_{\rm h}$, by approaching the
black hole horizon at the southern hemisphere. This trajectory also corresponds to the outgoing
photon for the black hole with an opposite direction of spin. The difference in color along the photon trajectory is used for better visualization and does not have any special meaning.} \label{fig2}
\end{figure}

Fig.~\ref{fig1} shows the numerically calculated ``plunging'' trajectory of the planet
infalling into the rotating black hole.

From equations of motion (\ref{phimot}) and (\ref{tmot}) follows the crucial feature of any
test particle trajectory, approaching the event horizon at $x=x_{\rm h}$. Namely, by
approaching to the black hole horizon, $x\to x_{\rm h}$, the trajectory is quasi-periodically
``winding up'' with an azimuthal angular velocity  $\Omega_\varphi=d\varphi/dt$, coming close
to the angular velocity of the black hole horizon $\Omega_{\rm h}$, where
\begin{equation}
 \Omega_{\rm h}=\frac{d\varphi}{dt}\Big|_{x\to x_{\rm h}}
 =\frac{a}{2(1+\sqrt{1-a^2})}.
 \label{OmegaH}
 \end{equation}
This general behavior of the plunging trajectories near the event horizon is valid even for the
massless particles. See in Fig.~\ref{fig2} the corresponding plunging photon trajectory,
quasi-periodically ``winding up'' with the angular velocity $\Omega_\varphi\to\Omega_{\rm h}$,
by approaching the black hole event horizon at the southern hemisphere. As a result, the
angular velocity of black hole event horizon $\Omega_\varphi$ must be inevitably imprinted to
the QPO signal from the accreting black holes.

Any source of radiation, e.\,g., clump of the hot plasma, approaching the event horizon of the
rotating black hole will be viewed by the distant observer in a relativistic ``synchrotron
mode'' as short splashes of radiation, beamed and boosted forward into the narrow solid angles
\cite{bpt72,Misner72,Cunningham72,Polnarev72}, and repeated quasi-periodically with a frequency
very near to $\nu_{\rm h}=\Omega_{\rm h}/2\pi$.

The oscillation with $\Omega_{\rm h}$ from (\ref{OmegaH}) is a first observational signature
for revealing the spin of the supermassive black hole in the Galactic Center. The corresponding
second signature is related with the QPOs of non-equatorial bound orbits in the accretion flow.

\section{Quasi-periodic bound orbits}

\begin{figure}[th]
\begin{center}
\includegraphics[angle=0,width=0.99\textwidth]{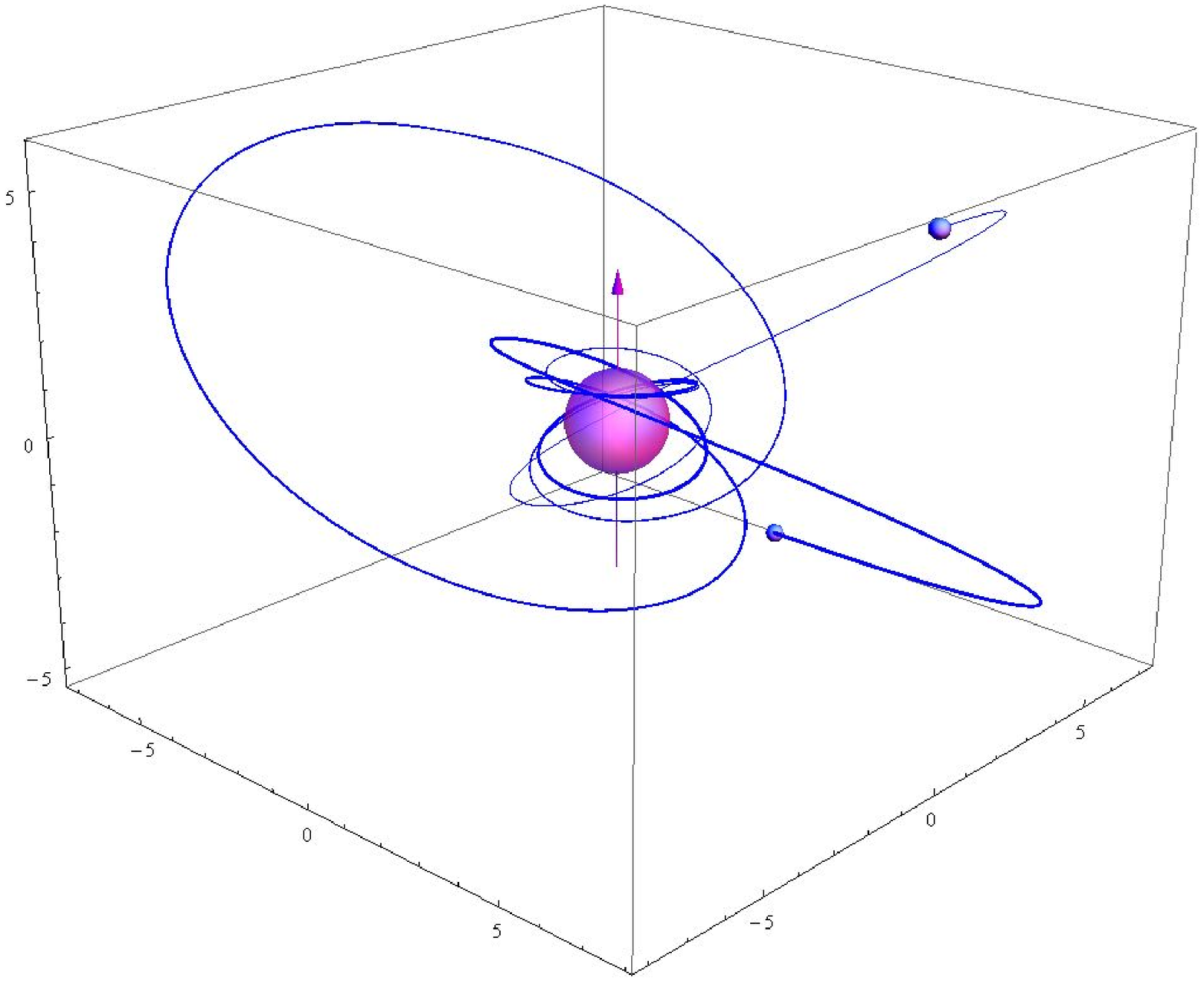}
\end{center}
\caption{The bound quasi-periodic orbit with $Q=2$, $E=0.92$, $L=1.9$, $x_{\rm p} =1.74$,
$x_{\rm a}=9.48$ and $\theta_{\rm max}=36.3^\circ$, viewed from the north pole of the black
hole with $a=0.9982$ and $x_{\rm h}=1.059$. The orbit is shown thin at the beginning and thick
at the ending.} \label{fig3}
\end{figure}
\begin{figure}[th]
\begin{center}
\includegraphics[angle=0,width=0.99\textwidth]{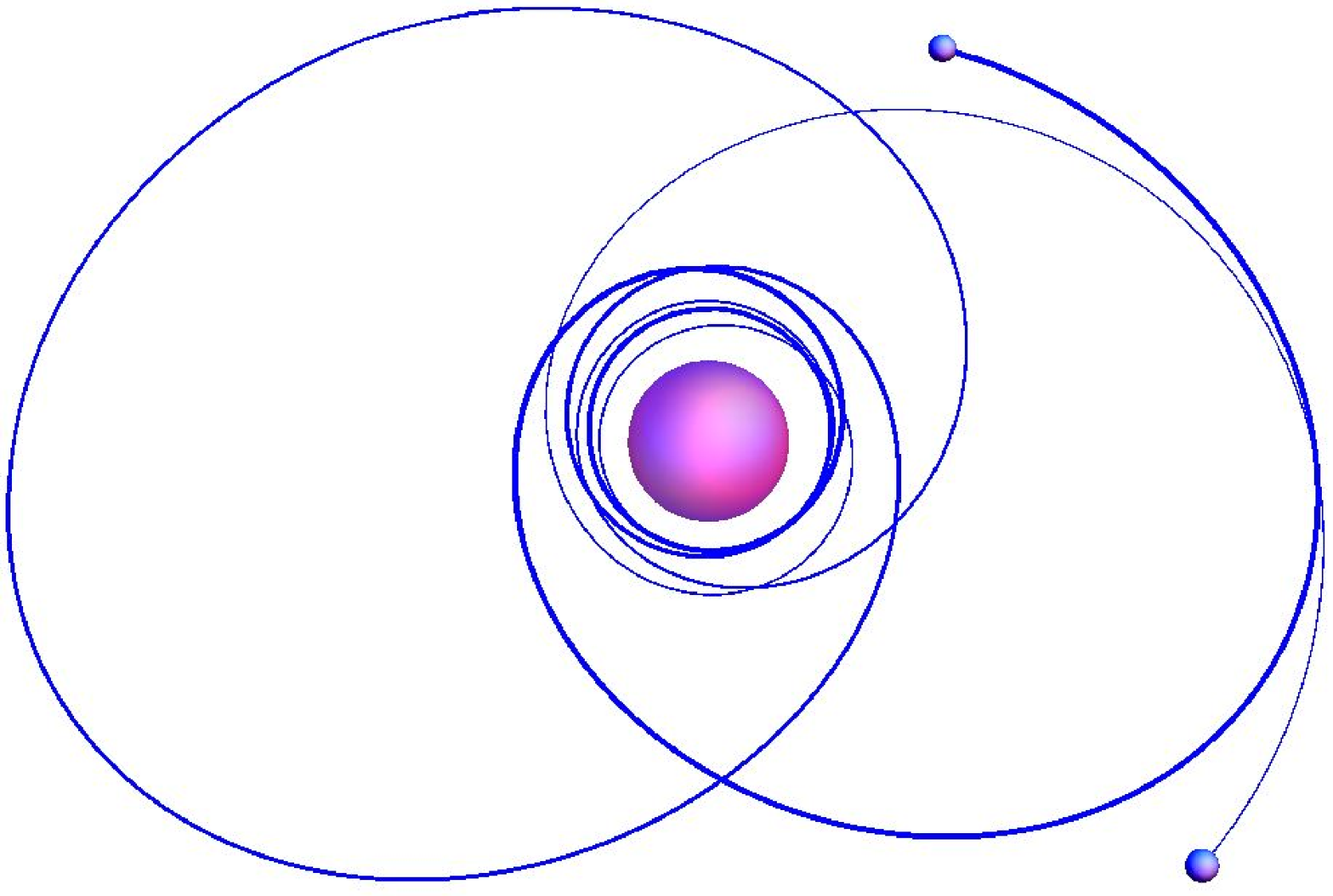}
\end{center}
\caption{The same orbit as in Fig~\ref{fig3}, viewed from the north pole of the black hole.}
\label{fig4}
\end{figure}

The specific features of the accretion disk flow near the black hole depend in part on the
properties of the stable gravitationally bound orbits of test particles with $E<1$
\cite{Wilkins72}, which have two radial turning points, the apsis and periapsis radii,
$r_{\rm a}$ and $r_{\rm p}$, respectively, and also two latitudinal turning points, $\pi/2\pm
\theta_{\rm max}$ above and below the equatorial plane, where $\theta_{\rm max}$ is a maximal
elevation angle of the orbit with respect to the equatorial plane. These turning points are
defined by the zeroes of the corresponding effective potentials $V(r)$ and $V(\theta)$ from
(\ref{Vr}) and (\ref{Vtheta}) respectively. The bound orbits around the rotating Kerr black
hole are quasi-periodic in general, because metric depends not on the one, but on the two
coordinates: the radius $r$ and the latitude $\theta$, in contrast to the Schwarzschild case.
The bound orbits oscillate not in time but only in space between the two radial turning points and between the two latitudinal turning points. The pure periodic bound orbits in the Kerr metric are only the degenerate ones: they either confined in the equatorial plane of the black hole or belong to the specific case of spherical orbits \cite{Wilkins72}, with a radial coordinate $r=const$.

\begin{figure}[ht]
\begin{center}
\includegraphics[angle=0,width=0.95\textwidth]{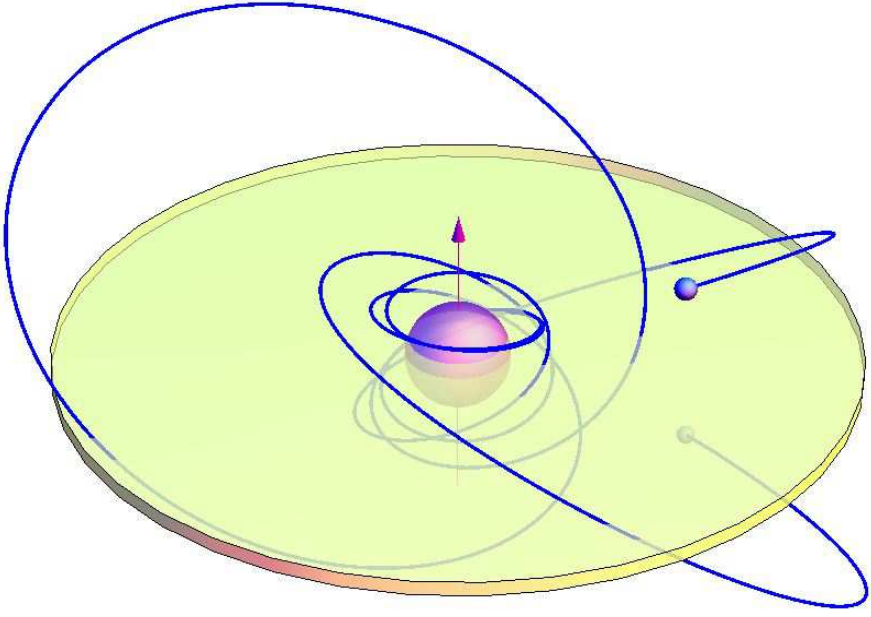}
\end{center}
\caption{The same bound quasi-periodic orbit as in Fig.~\ref{fig3} and \ref{fig4}, viewed
aside, and superimposed onto the thin and opaque accretion disk. Only the parts of the orbit,
which are above the accretion disk, may be viewed by the distant observer as QPOs with a
frequencies near $\nu_{\rm \theta}=\Omega_{\rm \theta}/2\pi$.} \label{fig5}
\end{figure}

See in Figs.~\ref{fig3} and \ref{fig4} an example of the numerically calculated bound
quasi-periodic orbit, viewed from the black hole north pole and aside, respectively. In the
Table~\ref{tbl-1} are shown the successive ($n=1,2,3\ldots$) radial ($T_{n,r}$, $\tau_{n,r}$),
azimuthal ($T_{n,\varphi}$, $\tau_{n,\varphi}$) and latitudinal periods  ($T_{n,\theta}$,
$\tau_{n,\theta}$) for the  bound quasi-periodic orbit, shown in Figs.~\ref{fig3} and
\ref{fig4} and measured by the distant ($T$) and proper ($\tau$) observers, respectively. In
Fig.~\ref{fig5} this orbit is superimposed onto the thin and opaque accretion disk, when only
the parts of the orbit, which are above the accretion disk, may be viewed by a distant
observer. This is an illustration of the QPOs produced by one of the numerous hot spots or
clumps of plasma in the accretion flow, which may be used for the determination of the black
hole mass and spin \cite{Syunyaev73,Abramowicz92,Zakharov94,Broderick06,Wang12}.

The presence of bright plasma points in the turbulent accretion flow are also confirmed by numerical simulations of the thin accretion disks \cite{Hawley01,Armitage01,Reynolds01}. Below we
calculate the corresponding oscillation periods of test particles (hot spots of plasma) in the
thin accretion disk.

Figs.~ \ref{fig3}, \ref{fig4} and \ref{fig5} demonstrate the qualitative features of the general non-equatorial orbits of test particles (planets or hot spots) around the Kerr black hole. It is supposed in the following that the observed QPO phenomena may be explained by the light curves of the hot spots on the nearly circular orbits.

\begin{table}
\begin{center}
\caption{Successive ($n=1,2,3\ldots$) radial, azimuthal, and latitudinal periods
($T_{n,r},T_{n,\varphi},T_{n,\theta}$) in units $MG/c^3$, measured by the distant observer, for
the  bound quasi-periodic orbit, shown in Figs.~\ref{fig3} and \ref{fig4}.  \label{tbl-1}}
\begin{tabular}{c|c|c|c} %{rrrrrr}
\hline\hline
$n$ & Azimuthal period & Radial period & Latitudinal period \\
  & $T_{n,\varphi}$ & $T_{n,r}$ & $T_{n,\theta}$ \\
\hline
1  & 74.2  & 181.1  & 90.4  \\
2  & 18.9  & 180.6  & 61.9  \\
3  & 22.4  &        & 108.6 \\
4  & 121.3 &        & 40.8  \\
5  & 26.3  &        &       \\
6  & 19.5  &        &       \\
7  & 44.3  &        &       \\
\hline
\end{tabular}
\end{center}
\end{table}

\section{Orbital oscillations}

The azimuthal and latitudinal angular velocities of the non-equatorial bound orbits,
$\Omega_\varphi$ and $\Omega_{\rm \theta}$, respectively, are related by the equation
\cite{Wilkins72,Merloni99}:
\begin{equation}
\frac{\Omega_{\rm \theta}}{\Omega_\varphi}\!=\!\frac{\pi}{2}(\beta z_+)^{1/2}
\!\!\left[\frac{L}{a}\Pi(-z_-,k)\!+\!\frac{2xE\!-\!aL}{\Delta}K(k)\right]^{-1}\!\!\!\!,
 \label{Omegathetaphi}
 \end{equation}
where and $K(k)$ and $\Pi(-z_-,k)$ are, respectively, the full elliptic integrals of the first
and third kind, $k^2=z_-/z_+$,
\begin{equation}
 z_{\pm}=(2\beta)^{-1}[\alpha+\beta\pm\sqrt{(\alpha+\beta)^2-4Q\beta}],
 \label{zpm}
 \end{equation}
$\alpha=(Q+L^2)/a^2$ and $\beta=1-E^2$. By using the values for orbital energy $E$ and
azimuthal angular momentum $L$ from \cite{dokuch11,dokuch12} for the test particle at the
non-equatorial spherical orbit, it can be calculated from equations of motion (\ref{phimot})
and (\ref{tmot}) the corresponding azimuthal angular velocity at the equatorial plane:
\begin{equation}
 \Omega_{\varphi,\rm sph}\!=\!\frac{x\sqrt{x^3(3Q\!-\!Qx\!+\!x^2)\!+\!a^2Q^2}\!-\!a(x^2\!+\!3Q)}
 {\{x^5-a^2[x^2+Q(x+3)]\}M}.
 \label{Omegaphisph}
 \end{equation}
In the particular case of circular orbits in the equatorial plane ($r=const$, $Q=0$,
$\theta=\pi/2$), the angular velocity (\ref{Omegaphisph}) is simplified to the well known form
\cite{bpt72}:
\begin{equation}
 \Omega_{\varphi,\rm circ}=\frac{1}{a+x^{3/2}}\frac{1}{M}.
 \label{Omegaphicirc}
 \end{equation}
From (\ref{Omegathetaphi}) and (\ref{Omegaphisph}) in the limit $Q\to0$ follows the angular
velocity of the latitudinal oscillation of a near circular orbit in the thin accretion disk:
\begin{equation}
\Omega_{\rm \theta}=\frac{2\pi}{T_{\rm \theta}}
=\frac{\sqrt{x^2-4ax^{1/2}+3a^2}}{x(a+x^{3/2})}\frac{1}{M}.
 \label{OmegathetaQ0}
 \end{equation}
This angular velocity describes the latitudinal oscillation of the hot spot or clump of plasma
in the thin accretion disk.

It is assumed that the brightest hot spots in the accretion flow are located near the inner
edge of the accretion disk, corresponding to the radius of the marginally stable circular orbit
\cite{bpt72}, $x=x_{\rm ms}$:
\begin{equation}
x_{\rm ms}=3+Z_2-\sqrt{(3-Z_1)(3+Z_1+2Z_2)},
 \label{Z12}
 \end{equation}
where
\begin{equation}
Z_1=1+(1-a^2)^{1/3}[(1+a)^{1/3}+(1-a)^{1/3}]
 \label{Z1}
 \end{equation}
and $Z_2=\sqrt{3a^2+Z_1^2}$.

\begin{figure}%[h]
\begin{center}
\includegraphics[angle=0,width=0.95\textwidth]{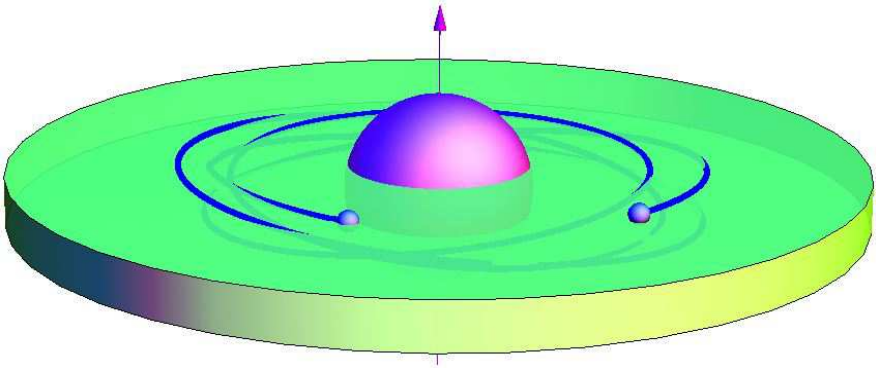}
\end{center}
\caption{Near the equatorial orbit of the bright clump of plasma with $Q=0.1$, $E=0.91$,
$L=2.715$, $x_{\rm p} =3.85$, $x_{\rm a}=5.01$ and $\theta_{\rm max}=6.6^\circ$, oscillating in
the thin opaque accretion disk around the black hole with $a=0.65$, $x_{\rm h}=1.76$ and
$x_{\rm ms}=3.62$.} \label{fig6}
\end{figure}

See in Fig.~\ref{fig6} the example of the oscillating clump of
plasma in the thin accretion disk around the moderately fast rotating black hole.

The latitudinal oscillation with an angular velocity $\Omega_{\rm \theta}$ from
(\ref{OmegathetaQ0}), estimated at the radius $x=x_{\rm ms}$, is the second requisite
observation signature of the spinning black hole in the Galactic Center.

\section{Black hole spin and mass in the Galactic Center}

\begin{figure}%[th]
\begin{center}
\includegraphics[angle=0,width=0.95\textwidth]{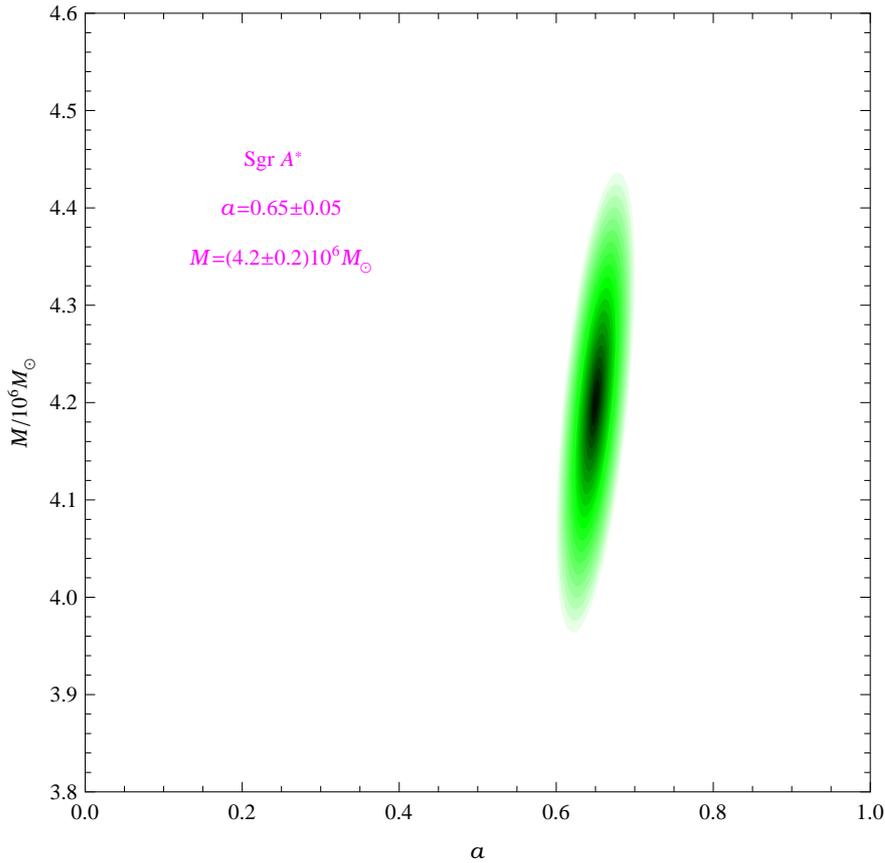}
\end{center}
\caption{The 1-$\sigma$ error ($M,a$)-region for the joint resolution of equations (\ref{OmegaH})
and (\ref{OmegathetaQ0}) with the observed 11.5 min QPO1 period, identified with $T_{\rm h}$,
and, respectively, the 19 min period QPO2, identified with $T_{\rm \theta}$.} \label{fig7}
\end{figure}

Fig.~\ref{fig7} shows the 1-$\sigma$ error ($M,a$)-region for the joint resolution of equations
(\ref{OmegaH}) and (\ref{OmegathetaQ0}) with the observed 11.5 min QPO1 period, identified with
$T_{\rm h}$, and, respectively, the 19 min period QPO2, identified with $T_{\rm \theta}$. This
1-$\sigma$ ($M,a$) region corresponds to the Kerr metric rotation parameter, $a=0.65\pm0.05$, and
mass, $M=(4.2\pm0.2)10^6M_\odot$, for the supermassive black hole in the Galactic Center.

\begin{figure}%[th]
\begin{center}
\includegraphics[angle=0,width=0.95\textwidth]{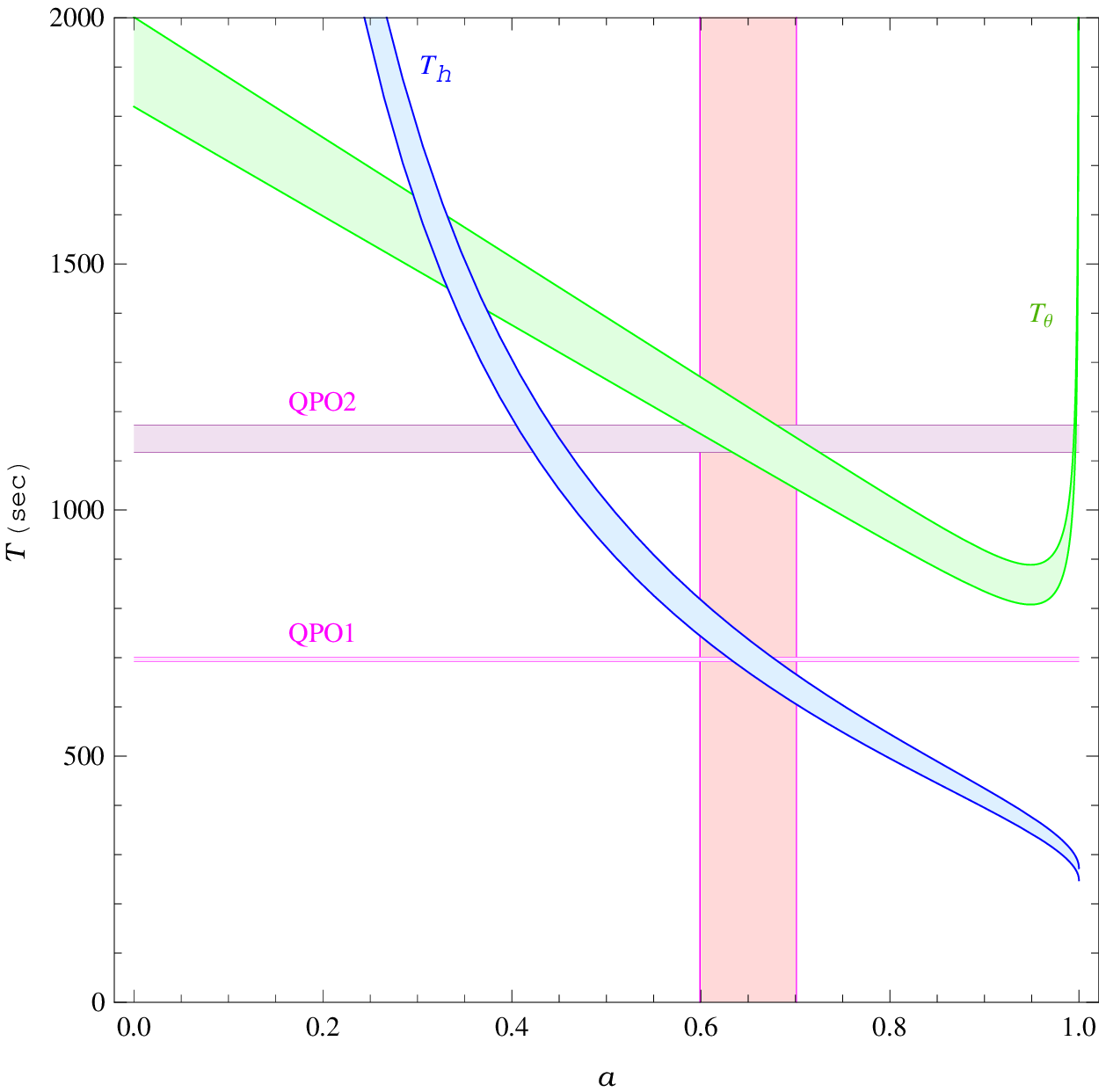}
\end{center}
\caption{The observed QPOs with the mean periods 11.5 and 19 min (filled horizontal stripes
QPO1 and QPO2) from the supermassive black hole Sgr~A*, identified, respectively, with a period
of the event horizon rotation $T_{\rm h}$ from (\ref{OmegaH}) and a period of the latitudinal
oscillation of the hot plasma clump at the near-circular orbit in the thin accretion disk with
$\Omega_{\rm \theta}$ from (\ref{OmegathetaQ0}). The filled region for $T_{\rm \theta}$
corresponds to the numerically calculated permissible values of $Q$ in (\ref{Omegathetaphi}) and (\ref{Omegaphisph}), adjusted with the observational errors of QPO1 and QPO2 periods. The joint resolution of equations (\ref{OmegaH}) and (\ref{OmegathetaQ0}) with the observed values of QPO periods $T_{\rm h}$ and $T_{\rm \theta}$ reveals the black hole spin, $a=0.65\pm0.05$ and mass, $M= (4.2\pm0.2)10^6M_\odot$, shown in
Fig~\ref{fig7}.} \label{fig8}
\end{figure}

It is clearly illustrated in Fig.~\ref{fig8}, that a self-consistency of the observed QPO periods
with $T_{\rm h}$ and $T_{\rm \theta}$ corresponds to the same value of the black hole spin,
$a=0.65\pm0.05$. At the same time the values of azimuthal angular velocities $\Omega_\varphi$
of hot spots in the accretion disk are spread in a wide range. For this reason the azimuthal
oscillations in the accretion disk would not produce any prominent features in the spectrum of
QPOs.

Note also, that the additional three QPOs observed in the X-rays with periods around $1.7$,
$3.6$ and $37.5$ min \cite{Aschenbach04}. These QPOs are less prominent than the used ones and
seemingly related with the resonances in the accretion disk
\cite{Genzel03,Aschenbach04,McClintockRemillard06,Kato10,Dolence12} or
being the harmonics of the used QPOs with periods $11.5$ and $19$ min, respectively.

\section{Conclusion}

Interpretation of the known QPO data by dint of signal modulation from the hot spots in the
accreting plasma reveals the Kerr metric rotation parameter, $a=0.65\pm0.05$, of the
supermassive black hole in the Galactic Center. At the same time, the observed 11.5 min QPO
period is identified with the period of the black hole horizon rotation, and, respectively, the
19 min period is identified with the latitudinal oscillation period of hot spots in the
accretion flow. A major part of uncertainty in estimation of the black hole spin related with
an error in the measurements of the black hole mass in the Galactic Center.

A supermassive black hole in the Galactic Center acquires its angular momentum by accretion of
tidally destructed stars and gas clouds with accidentally orientated individual angular
momenta. For this reason a moderate spin value of the supermassive black hole Sgr A* is quite
natural due to specific conditions in the Galactic Center. Note also that the value of spin
parameter $a=0.65\pm0.05$, derived here by dint of QPOs, is in a qualitative agreement with the
corresponding quite independent estimation, $a\simeq0-0.6$, from the millimeter VLBI
observations of Sgr A* \cite{Broderick09,Broderick11}. At the same time, the black hole mass,
$M=(4.2\pm0.2)10^6M_\odot$, derived here from the QPOs, is in a good agrement with a quite
independent estimation, $M=(4.1\pm0.4)\,10^6M_\odot$, measured from observations of the fast
moving S0 stars \cite{Gillessen09,Aschenbach04}.

\begin{acknowledgements}
I thank V. A. Berezin and Yu. N. Eroshenko for helpful discussions. This research was supported in part by the Russian Foundation for Basic Research grant RFBR~13-02-00257 and by the Research Program OFN-17 of the Division of Physics, Russian Academy of Sciences.
\end{acknowledgements}

\end{document}